\documentclass[12pt,amssymb,epsfig]{article}
\usepackage{latexsym}
\usepackage{amsfonts}
\usepackage{amsmath}
\usepackage{amssymb}
\usepackage[dvips]{graphicx}
\usepackage{makeidx}
\usepackage{epsfig}
\input  epsf
\newcommand{\anticom}[2]{\ensuremath{\left \{ #1 , #2 \right\}}}
\newcommand{\com}[2]{\ensuremath{\left [ #1 , #2 \right]}}
\def\rlx{\relax\leavevmode}
\def\inbar{\vrule height1.5ex width.4pt depth0pt}
\def\IZ{\rlx\hbox{\small \sf Z\kern-.4em Z}}
\def\IR{\rlx\hbox{\rm I\kern-.18em R}}
\def\ID{\rlx\hbox{\rm I\kern-.18em D}}
\def\IC{\rlx\hbox{\,$\inbar\kern-.3em{\rm C}$}}
\def\zero{\rlx\hbox{\,$\inbar\kern-.18em{ 0}$}}
\def\IN{\rlx\hbox{\rm I\kern-.18em N}}
\def\one{\hbox{{1}\kern-.25em\hbox{l}}}
\def\beq{\begin{equation}}
\def\eeq{\end{equation}}
\def\bea{\begin{eqnarray}}
\def\eea{\end{eqnarray}}
\def\ber{\begin{array}}
\def\eer{\end{array}}

\begin{document}

\begin{titlepage}

July 2002 \hfill{UTAS-PHYS-02-02}\\
% \mbox{}\hfill {CMA-01-yy}
% \mbox{} \hfill{hep-th/yymmdd}
\vskip 1.6in
\begin{center}
{\Large {\bf Polar decomposition of a
Dirac spinor}}\end{center}

\normalsize
\vskip .4in

\begin{center}
J G Sumner and P D Jarvis
\par \vskip .1in \noindent
{\it School of Mathematics and Physics, University of Tasmania}\\
{\it GPO Box 252-21, Hobart Tas 7001, Australia }\\

\end{center}
\par \vskip .3in

\begin{center}
{\large {\bf Abstract}}\\
\end{center}
Local decompositions of a Dirac spinor into `charged' and `real'
pieces $\psi(x) = M(x) \chi(x)$ are considered. $\chi(x)$ is a
Majorana spinor, and $M(x)$ a suitable Dirac-algebra valued field.
Specific examples of the decomposition in $2+1$ dimensions are
developed, along with kinematical implications, and
constraints on the component fields within $M(x)$ sufficient to
encompass the correct degree of freedom count. Overall local
reparametrisation and electromagnetic phase invariances are
identified, and a dynamical framework of nonabelian gauge theories
of noncompact groups is proposed. Connections with supersymmetric
composite models are noted (including, for 2+1 dimensions,
infrared effective theories of spin-charge separation in models of
high-$T_{c}$ superconductivity).

%%%%%%%%%%%%%%%%%%%%%%%%%%%%%%%%%%%%%%%%%%%%%%%%%%%%%%%%%%%%%%%%%%%%%%%%%%%%%%%%%%%%%%%%%%%%%
%                                                                                           %
%   \vfill                                                                                  %
%   \noindent                                                                               %
%   \footnotesize $^{*}$Centre for , Uni  of                                               %
%   \texttt{name@server.domain.edu.au}                                                   %
%                                                                                           %
%%%%%%%%%%%%%%%%%%%%%%%%%%%%%%%%%%%%%%%%%%%%%%%%%%%%%%%%%%%%%%%%%%%%%%%%%%%%%%%%%%%%%%%%%%%%%
\end{titlepage}

%\pagebreak
\section{Introduction}
The standard model of elementary particle physics has been highly
successful in accounting for the structure and interactions of
the currently known generations of fundamental quarks and leptons
and associated charge carrying gauge bosons. However, there is
continued interest in alternative formulations, in the hope of
reducing the arbitrariness and hierarchical fixing of standard
model parameters, seeking physics beyond the standard model, or
perhaps in improving the singular mathematical structure of
quantum field theory. Supersymmetric models, composite models and
string-theory models have all been investigated under these
umbrellas.

A recurring theme in investigations of alternative formulations
has been the nature of spinorial quantities, given their
double-valuedness and the fact that physical quantities are always
even functions of them.  Frameworks purporting to eliminate the
`phase' of the spinor wavefunction have included K\"{a}hler
fermions \cite{Benn}, the Lounesto real Clifford algebra
\cite{Lounesto} and the Hestenes geometric algebra constructions
\cite{Hest}, and various attempts to substitute the four complex
components of a Dirac spinor in four dimensions (aside from the
unobservable overall phase), with seven bosonic, gauge invariant
quantities (usually defining a Lorentz frame, together with a
scalar field) \cite{Tak,mik,Rudolph}. In the present note we
take these ideas in a related direction, in proposing a different
rearrangement of the degrees of freedom of a Dirac spinor
wavefunction, which may entail insights into new dynamical models
of fermions.  We study `polar' decompositions of a Dirac spinor,
into a `charged' part and a `neutral' part, in the form of a local
product $\psi(x) = M(x)\cdot \chi(x) $.  Taking into account
charge conjugation properties, as well as Lorentz covariance,
$\psi'(x) = M'(x)\cdot \chi'(x) $ $= S(\Lambda) M(x)
S(\Lambda)^{-1} \cdot S(\Lambda) \chi(x)$ leads to the
\textit{Ansatz} that $\chi(x)$ be a \textit{Majorana} spinor (in
space-time dimensions $D$ and spinor dimensions $N$ where this is
defined), and that $M(x)$ be an appropriate Dirac-algebra valued
quantity (within the space ${\mathbb M}_{N}({\mathbb C})$ of $N
\times N$ complex matrices) transforming in the adjoint
representation.

In order to have a specific example to study we take the case of
$D=3$ ($2+1$-dimensional Minkowski space), with spinor dimension
$N=2$.  In the next section (\S 2) we argue for a particular
standard form of the parametrisation, and by studying aspects of
this are able to demonstrate the decomposition, the number and
nature of the parameters and electromagnetic interactions.
Particular examples of the free particle solutions of the Dirac
equation are given in the appendix, \S A, for both plane and
circular waves. In \S 3 below, we return to the general
parametrisation from the point of view of local gauge symmetries
and suggest a dynamical picture of the polar decomposition.  In
the concluding remarks (\S 4 below) the case of general space-time
dimensions is discussed, together with implications of the polar
decomposition in the setting of supersymmetric models,
applications in condensed matter theory, as well as the context of
nonabelian gauge theory.

\section{Polar decomposition in $2\!+\!1$ dimensions}
For completeness, a summary of notational conventions relating to the
Dirac algebra in $2+1$ space-time dimensions is given in the
appendix. The most important observation is that the set of $2 \times
2$ complex matrices (to which the `charged' part $M(x)$ of the Dirac
wavefunction belongs) is spanned by the four matrices $\one$,
$\gamma^{\mu}$, $\mu = 0,1,2$.
Thus we begin by assuming a decomposition of the form of a product
of a bosonic field and a (Majorana) fermionic field,
\begin{align}\label{covariant}
\psi(x) = M(x) \chi(x) = \left[a(x)
+b_{\mu}(x)\gamma^{\mu}\right]\chi(x),
\end{align}
where $a$ and $b_{\mu}$ are real.  For now, this can be interpreted as
a general parametrisation, in that it is the sum of a Majorana spinor
$a\chi$ and an anti-Majorana spinor $b_{\mu}\gamma^{\mu}\chi$ (the
general case is discussed in \S 4 below).  Clearly the parts transform
under the Lorentz group variously as a scalar, a vector and a spinor
field.  This decomposition has too many parameters if it is to
represent the four real degrees of freedom of a Dirac spinor.  The
scalar $a$ can always be scaled out, but $b_{\mu}$ is a three vector,
so there must be a condition which will reduce the number of free
parameters to 2.  Fortunately, requiring form invariance under local
$U(1)$ gauge transformations leads to such a condition.  These
transformations are given in terms of $\psi$ and the electromagnetic
vector potential by
\begin{align}
A_{\mu}\rightarrow{A_{\mu}}'&=A_{\mu}-\partial_{\mu}\alpha(x),\nonumber\\
\psi\rightarrow{\psi}'&=e^{iq\alpha(x)}\psi.\nonumber
\end{align}

The situation is complicated by the fact that the components of
the polar decomposition get mixed up in order to absorb the
phase. The appropriate procedure is to split the decomposition
into Majorana and anti-Majorana parts, apply the gauge
transformation and then collect respective Majorana and
anti-Majorana parts back together. This leads to
\begin{align}
\psi'=e^{iq\alpha}\psi=(a\cos{q\alpha}+ib_{\mu}\gamma^{\mu}\sin{q\alpha})\chi+(b_{\mu}\gamma^{\mu}\cos{q\alpha}+ia\sin{q\alpha})\chi.\nonumber
\end{align}
The above implies the transformations
\begin{align}
a\chi\rightarrow(a\chi)'&=\left(a\cos{q\alpha}+ib_{\mu}\gamma^{\mu}\sin{q\alpha}\right)\chi\nonumber\\
b_{\mu}\gamma^{\mu}\chi\rightarrow\left({b}_{\mu}\gamma^{\mu}\chi\right)'
&\equiv\frac{b'_{\mu}\gamma^{\mu}}{a'}\left(a\cos{q\alpha}+ib_{\mu}\gamma^{\mu}\sin{q\alpha}\right)\chi\nonumber\\
&=(b_{\mu}\gamma^{\mu}\cos{q\alpha}+ia\sin{q\alpha})\chi.\nonumber
\end{align}
A necessary condition for a solution is
\begin{align}
\frac{b'_{\mu}\gamma^{\mu}}{a'}
&=\frac{(b_{\mu}\gamma^{\mu}\cos{q\alpha}+ia\sin{q\alpha})(a\cos{q\alpha}-ib_{\mu}\gamma^{\mu}\sin{q\alpha})}{a^2\cos^2{q\alpha}+b^2\sin^2{q\alpha}}\nonumber\\
&=\frac{ab_{\mu}\gamma^{\mu}+i\sin{q\alpha}\cos{q\alpha}(a^2-b^{2})}{a^2\cos^{2}{q\alpha}+b^{2}\sin^{2}{q\alpha}}.\nonumber
\end{align}
Thus under gauge transformations, the vector part of the assumed
decomposition has covariance problems, unless
$a^2=b_{\mu}b^{\mu}$. Take this as a constraint, so that the
polar decomposition can be written (after choosing without loss of
generality $a >  0$ for nonzero $\psi(x)$),
\begin{align}\label{finalform}
\psi(x) = M(x) \chi(x), \qquad M(x) = \frac
12\left[1+n_{\mu}(x)\gamma^{\mu}\right], \qquad n_{\mu}n^{\mu} = 1
\end{align}
where $a$ has been scaled out.  In this form the polar
decomposition takes on the form of a true Dirac spinor, with both
Majorana and anti-Majorana parts.

The decomposition now has a particularly elegant behaviour under
gauge transformations:
\begin{align}
\chi\rightarrow\chi'&=\exp\left({iqn_{\beta}\gamma^{\beta}\alpha}\right)\chi,\nonumber\\
n_{\mu}\rightarrow{n}'_{\mu}&=n_{\mu}.
\end{align}
So not only has form invariance under a gauge transformation been
achieved, but a new restriction on the decomposition has been
found so that the correct number of free parameters is present.
Notice also that the gauge transformation is in fact a
\textit{local} Lorentz transformation on the Majorana spinor (see
(\ref{spinrep})).

Let us consider the implications of (\ref{finalform}) for charge
conjugation, $\psi \rightarrow \psi_{c}$. We have
\begin{align}
\psi\rightarrow\psi_c =M_{c}\chi =
\textstyle{\frac{1}{2}}(1-n_{\mu}\gamma^{\mu})\chi.
\end{align}
Thus, the bosonic part of the wavefunction $n_{\mu}$ goes to
$-n_{\mu}$ under charge conjugation\footnote {The apparent gauge
invariance of $n_{\mu}$ in this parametrisation is the result of
additional reparametrisation invariance (see below).}. This is
consistent with the observation that the bosonic part of the
wavefunction carries the interaction with the electromagnetic
field. The constraint $n_{\mu}n^{\mu}=1$ entails
\begin{align}
(n_0)^2=1+\mathbf{n}.\mathbf{n},\nonumber
\end{align}
so that the parameter space of $n_0$ is the two disconnected
regions $n_0>1$ and $n_0<1$, corresponding to a two-sheeted
hyperboloid. Thus it is actually the sign of $n_0$ alone that
characterises the charge conjugation symmetry of the wavefunction.

An expression for $\chi$ is calculated easily as
\begin{align}
\chi=\textstyle{\frac{1}{2}}(\psi+\psi_c);
\end{align}
the Majorana part of the wavefunction is simply the superposition
of the original spinor with its charge conjugate.
Consider the scalar and
vector covariants
\begin{alignat} {2}
\overline{\psi}\psi&=\textstyle{\frac{1}{2}}\overline{\chi}\chi,
\qquad
\overline{\psi}\gamma^{\mu}\psi=\textstyle{\frac{1}{2}}n^{\mu}\overline{\chi}\chi.\nonumber
\end{alignat}
Hence the scalar density of the Dirac field is carried exclusively by
the Majorana-spinor component of the field, while the Noether current
of the Dirac field is represented by $n_{\mu}$ scaled by the scalar
density $\overline{\chi}\chi$.  A remarkably simple form for
$n_{\mu}$ follows as
\begin{align}\label{soln}
n^{\mu}=\frac{\overline{\psi}\gamma^{\mu}\psi}{\overline{\psi}\psi}=
\frac{j^{\mu}}{\overline{\psi}\psi},
\end{align}
so that $n^{\mu}$ can be interpreted as the 3-velocity of
the Dirac field. 

Finally, notice that both $M$ and its charge conjugate $M_{c} = \frac
12(1-n_{\mu}\gamma^{\mu})$ are singular,
\begin{align}\label{dets}
\det{M}=\det{M}_c= \frac 14( 1- n_{\mu}n^{\mu})=0.
\end{align}
This seems to imply that if we take
$\psi=M\chi$ then $\chi$ is not unique. However, (\ref{dets}) is
equivalent to
\begin{alignat}{3}
M^2=M,\qquad{M}^2_c=M_c,\qquad{M}M_c=0,\nonumber
\end{alignat}
showing that $M$ and $M_c$ are orthogonal projectors.  The null
space of $M$ is then any spinor written in the form $M_c\varphi$.
For the decomposition this means that
\begin{align}
\psi&=M\chi\nonumber\\
&=M(\chi+M_c\varphi).\nonumber
\end{align}
However it is clear that $M_c\varphi$ cannot be Majorana so that
$\chi$ is in fact unique. Notice also that $\psi$ is orthogonal to
its charge conjugate:
\begin{align}
\frac{1}{2}\psi^*_{1}\psi^*_{2} =
\overline{\psi}\psi_c=\overline{{\chi}M}M_c\chi=\overline{\chi}MM_c\chi=
\chi_{2}\chi_{1}\det{M} = 0,
\end{align}
due to the projective property of $M$; similarly the complex vector 
$\overline{\psi}\gamma^{\mu}\psi_c$ vanishes, for real $n^{\mu}$.

\section{Local gauge symmetries}

In the previous section the discussion focussed on choosing a
minimally constrained set of parameters for the polar-type
decomposition in the $2+1$-dimensional case.  In this section we
reconsider the general \textit{Ansatz} $\psi = M \cdot \chi$ from
the point of view of local gauge symmetries which carry redundant
parameters, and which may also play a dynamical role. Namely, the
product decomposition admits any \textit{local} reparametrisations
which respect the form of $\chi$ as a Majorana spinor. Given
$\chi$ as in (\ref{majcon}), the requirement that $L(x)\chi$ be of
the same form necessitates
$$
L(x) = \left( \begin{array}{cc} P & Q \\ Q^{*} & P^{*}
\end{array} \right), \quad P P^{*} - Q Q^{*} \ne 0, \quad
P,Q\epsilon\mathcal{C}.
$$
-- identical in this case to the pseudo-unitary transformations
acting on bispinors $\psi$ as in (\ref{spinrep}), preserving the
(Dirac) norm $\overline{\psi} \psi = \psi^{\dagger} \gamma^{0}
\psi = u^{*}u - v^{*}v$ up to proportionality. These are nothing
but \textit{local} Lorentz transformations, in fact, elements of
the covering group $SU(1,1)$ composed with local field  rescalings
belonging to the multiplicative group of dilatations, ${\mathbb R} - { \{ } 0 { \}
}  \simeq GL(1, {\mathbb R})$. Thus with our assumption that
$\chi$ is a Majorana spinor, we have the following semidirect
product of global Lorentz transformations $\Lambda \in SO(2,1)$
(in the spinor representation $S(\Lambda)$), local Weyl group
reparametrisations $L \in SU(1,1) \times GL(1, {\mathbb R})$ and
local electromagnetic phase transformations:
\begin{eqnarray*}
    M \rightarrow S(\Lambda) M S(\Lambda)^{-1}, &\quad& \chi \rightarrow S(\Lambda) \chi,\\
    M  \rightarrow  M L(x)^{-1} , &\quad& \chi \rightarrow   L(x) \chi, \\
      M \rightarrow e^{i\theta}M, &\quad&   \chi  \rightarrow  \chi.
\end{eqnarray*}

From this point of view the minimal parameters for $M$ of the
previous section can be seen as an appropriate gauge choice with
respect to the local Weyl group. Generically, the independent
degrees of freedom of $M$ are associated with the orbit structure
of ${\mathbb M}_{2}({\mathbb C})$ under the right action of $L$,
and the minimal parameters for $M$ correspond to a choice of
section of ${\mathbb M}_{2}({\mathbb C})$ under the induced
fibration. For example, the apparent \textit{non}-invariance of
$\chi$ under local phase transformations of $\psi$ --
contradictory if $\chi$ is a neutral field -- is the result of a
combined local phase transformation and local Weyl (actually
$SU(1,1)$) transformation in order to remain within the minimal
gauge class:
\[
e^{iq \theta}M \chi = M e^{iq \theta n^{\mu}\gamma_{\mu}} \chi
                    = M (e^{iq \theta n^{\mu}\gamma_{\mu}} \chi)
\]
where the projection property of $M = \frac 12 (1 +
n^{\mu}\gamma_{\mu})$ with $n\! \cdot\! n =1$ has been used, to
compensate the local phase transformation with a local Lorentz
transform with parameter $q\theta(x)n^{\mu}$, so that effectively
$M' = M$ and $\chi' = e^{iq \theta n^{\mu}\gamma_{\mu}} \chi$ as
discussed above.

The orbits of $M_{2}({\mathbb C})$ are most easily studied via the
unitary change of basis
\begin{eqnarray}
    \chi \rightarrow \widetilde{\chi} = \Theta \chi,
    &\quad& M \rightarrow \widetilde{M} = \Theta M \Theta^{-1},\nonumber \\
    \mbox{where} \quad \Theta = \textstyle{\frac 12} \left( \begin{array}{cc} 1 &  1
    \\ -i &  i \end{array} \right), &\quad&
    \mbox{so} \quad \widetilde{\chi} = \left( \begin{array}{c} r \\ s
    \end{array} \right),
    \label{eq:RealBasisDef}
\end{eqnarray}
with $u = r + i s$ in terms of the original components of $\chi$.
In this real basis\footnote{The Dirac matrices are pure
imaginary in this basis.} the Lorentz group clearly acts as $2
\times 2$ nonsingular real matrices (the group $SL(2, {\mathbb
R})$), with the local Weyl group being $GL(2, {\mathbb R})$.
Under $M \rightarrow M' = M \cdot L(x)$, $\mbox{det}M' =
\mbox{det}M \cdot \mbox{det}L$, so that orbits can be
classified\footnote{The determinants of the real and imaginary
parts of $M$ also scale with $\mbox{det}L(x)$.} according to
whether $\mbox{det}M =0$ or $\mbox{det}M \ne 0$.

Generically let
\[
\widetilde{M} = \left( \begin{array}{cc} a+i\alpha & b+i\beta \\
c+i\gamma & d+i\delta \end{array} \right)
\]
(corresponding to an expansion $M = A + B^{\mu}\gamma_{\mu}$ in
the Dirac algebra, with complex coefficients). Then $\mbox{det}M =
\mbox{det}\widetilde{M}= (\Delta - \Delta') +i (\Gamma + \Gamma')$
with $\Delta = ad-bc$, $\Delta' = \alpha \delta - \beta \gamma$,
$\Gamma = a \delta - b \gamma$, $\Gamma' = \alpha d - \beta c$.
Consider the case $\Delta \ne 0$, $\mbox{det}M \ne 0$. Let
$\widetilde{M} = e^{i \phi}\widetilde{N}$ where
$\mbox{det}\widetilde{N}$ is real, and set
$$
L(x) = \frac{1}{\Delta}\left( \begin{array}{cc} \; d & -b \\
-c & \; a \end{array} \right).
$$
Transforming back to the standard basis, we find for some
$b^{\mu}$
\begin{eqnarray}
    M' &=& e^{i \phi}(1 + b^{\mu}\gamma_{\mu}), \quad \mbox{det}M' =
    e^{2i\phi}(1 - b \!\cdot \! b )\ne 0.
    \label{eq:DetMne0}
\end{eqnarray}
On the other hand in the case $\Delta \ne 0$, $\mbox{det}M = 0$
the same choice of $L(x)$ yields
\begin{eqnarray}
    M' &=& (1 + n^{\mu}\gamma_{\mu}), \quad \mbox{det}M' =
    1 - n \!\cdot \!n = 0.
    \label{eq:DetMeq0}
\end{eqnarray}
Clearly the latter case (\ref{eq:DetMeq0}) coincides with the
minimal parametrisation of the previous section; the \textit{two}
real constraints $\mbox{det}M = 0$ entail $8-2=6$ parameters, or
effectively $6-4=2$ degrees of freedom resident in $M$, after
allowing for the gauge transformations by elements of
$GL(2,{\mathbb R})$ to fix a further four real components.  If
$\mbox{det}M$ is real but nonzero, there are \textit{three} field
components corresponding to the parametrisation $M = (a +
b^{\mu}\gamma_{\mu})$ for real $a$, $b^{\mu}$ constrained to some
hyperboloid, say $a^{2} - b\cdot b = 1$.  Finally in case
(\ref{eq:DetMne0}) above there are four fields $\phi$, $b^{\mu}$.

The surfeit of allowed parameters of $M$ in the last two cases
would be expected to be reconciled in a dynamical picture. For
example, the role of the four-dimensional inner product $a^{2} -
b\!\cdot \!b$ in the above discussion perhaps points to a higher
conformal symmetry formulation of these bosonic concomitants of
the Dirac spinor field (a projective null cone has codimension
two, effecting the desired reduction from four to two degrees of
freedom). An obvious first step is to promote the Weyl
reparametrisation freedom of the polar decomposition to a local
gauge principle, by introducing gauge fields $\omega_{\mu}^{ab}$
in the internal Lorentz algebra, together with scale gauge fields
$\varphi_{\mu}$, and covariant derivatives
\begin{eqnarray*}
\nabla_{\mu} \chi &=& \partial_{\mu} \chi + \textstyle{\frac 12}
\omega_{\mu}^{ab}\sigma_{ab}\chi + \varphi_{\mu}\chi \\
\nabla_{\mu} M &=& \partial_{\mu} M - \textstyle{\frac 12}
\omega_{\mu}^{ab}M\sigma_{ab} - \varphi_{\mu} M.
\end{eqnarray*}
Furthermore, we anticipate a total covariant derivative including
electromagnetic potential
$$
{\mathcal D}_{\mu} = \nabla_{\mu} + iq A_{\mu}
$$
so that the ordinary covariant derivative of the Dirac spinor
$\psi = M \cdot \chi$ becomes
$$
(i \gamma^{\mu}D_{\mu} + m) M \cdot \chi =
    \left[(i \gamma^{\mu}{\mathcal D}_{\mu} + m )M\right]\cdot \chi  +
               M \cdot (i \gamma^{\mu}\nabla_{\mu} \chi )  + {[}i
               \gamma^{\mu}, M {]} \partial_{\mu} \chi.
$$
Regarding $M$ and $\chi$ as the fundamental fields, it is natural
to interpret the Dirac equation for $\psi$ as the result of
separate `free' equations
\begin{equation}
(i \gamma^{\mu}{\mathcal D}_{\mu} + m) M = 0, \quad i
\gamma^{\mu}\nabla_{\mu} \chi =0 \label{eq:FreeMchieqns}
\end{equation}
plus coupling or `interaction' terms, which taken together would
re-constitute the third, commutator, piece of the above expansion
(which is not locally Weyl covariant as it stands).

\section{Conclusions}

In this note we have followed through, for the particular case of
2+1 dimensions, a proposal to re-interpret the degrees of freedom
of Dirac spinors in terms of `polar' type decompositions,
involving a neutral, Majorana part, and a charged companion.  Here
we have simply considered the kinematical implications of the
decompositions.  For the present case, the two degrees of freedom
of the Majorana spinor are accompanied by a (possibly complex)
scalar and vector field, which encapsulate the dynamics of a
further two degrees of freedom.  Under additional conditions,
these fields may be constrained, for example to the surface of a
hyperboloid or null cone in higher dimensions.  As an aside notice
that (\ref{finalform}) implies that there is no sense, under this
scheme, in which a generic Dirac spinor can be approximated as
some kind `charge perturbation' of a Majorana spinor.  This is
because the Majorana and anti-Majorana parts of the wavefunction
are of the same order of magnitude under all circumstances.
Therefore, from the point of view of classical equations of
motion, solutions of the Dirac equation under minimal coupling
with the electromagnetic field cannot be treated as perturbations
to Majorana solutions of the free equation.

In $3+1$ spacetime dimensions the polar decomposition $\psi = M
\cdot \chi$ of a Dirac spinor involves, apart from a Majorana
spinor (4 real degrees of freedom), the orbit structure of
$M_{4}({\mathbb C})$ under the local Weyl group $GL(4,{\mathbb
R})$, which can be expected to fix at most $32-16 = 16$ real
parameters.  The dynamics of the remaining component fields and
local gauge fields must therefore reduce the additional number of
degrees of freedom resident in $M$ from 16 to 4.

Further work to develop a concrete dynamical picture is required.
The connection fields implementing local Weyl reparametrisation
invariance parametrise a \textit{non-compact} gauge algebra.
Standard Yang-Mills actions are inconsistent (because the Killing
form occurring for terms quadratic in the field strength in the
action is non-positive definite for hermitean fields), and appeal
must therefore be made to other formulations\cite{Cahill}.
Alternatively, the analogy between the connection forms
${\omega^{ab}}_{\mu}$ and $\varphi_{\mu}$ and the vierbein
formulation of relativity (where local Lorentz invariance occurs
in the representation of the metric tensor) might be exploited,
so that Einstein-type actions for these fields may be possible.

Composite models of quarks and leptons commonly involve either
binding of multiparticle sub-quark or sub-lepton states by a
hypercolour gauge confinement mechanismm \cite{Volkas}, or
two-body bound states of scalars and fundamental
\textit{Ur-}fermions. Our proposal is evidently consistent with
the latter scenario, but motivated by a deconstruction of the
existing degrees of freedom of Dirac fields, rather than invoking
mere repetition of standard degrees of freedom at the preonic
level. Thus the role of the Majorana particles is to carry spin
degrees of freedom, while charge attributes are conferred by the
quantum numbers of the bosonic binding partners.  Whether
`confinement' exists for this type of model is unknown.

As noted above, our gauge bosons belong to a non-compact gauge
symmetry, and standard Yang-Mills theory cannot directly be
applied. What the degree of freedom count does guarantee is a
balance between fermionic and bosonic parts for each Dirac field,
the hallmark of possible supersymmetric formulations. In any case,
the existence of fermion generations is attributed, in standard
preon model terms, to the bound state spectrum. For example, if
(\ref{eq:FreeMchieqns}) are regarded as `free' equations of
motion, different fermion generations may be associated with zero
modes of a covariant Dirac-type operator, which then bind with the
charged scalar. In turn, the equation of motion of the latter
acquires curvature terms (in the noncompact gauge symmetry) which
affect its mass, and hence ultimately the mass of known quarks and
leptons.

There is a striking analogy between the foregoing considerations
(for relativistic theories) and the known situation in the
condensed matter context, again in $2+1$ dimensions, in which
spin-charge separation occurs into `spinon' and `holon' degrees of
freedom.  Relativistic formulations are appropriate at particular
nodes of the Fermi surface, and indeed `supersymmetric composite
models' have been advocated to describe an effective infrared
field theory of the phenomena of high-$T_{c}$ superconductivity,
\cite{sarkar}. Again, the implication for the relativistic context
would be that there is indeed a phase of matter wherein
fundamental fermions fractionate into their constituents, with the
local gauge fields taking on the dynamical role played by the
lattice.  In 2+1 dimensions the ingredients identified in this
work (Majorana spinor, and bosonic accompaniments accounting for 2
degrees of freedom) are necessarily the same as for matter
superfields (for $N=1$ supersymmetry).  The analogy with
supersymmetry is even closer in terms of the specific fields
$a(x)$, $b_{\mu}(x)$ parameterizing the matrix $M(x)$, if, as the
result of algebraic equations of motion and for appropriate gauge
fixing conditions, constraints such as $b_{\mu} = \partial_{\mu}
a$ (for complex $a$, $b_{\mu}$) arise. Further development of
our model would involve consideration of possible supersymmetric
formulations \cite{nahm}, and dynamical aspects.

In the case of the `polar' decomposition $\psi = M \cdot \chi$ of
Dirac spinors in a nonabelian gauge theory, it is natural to
adopt a generalised \textit{Ansatz} in which quantum numbers are
shared between $M$ and $\chi$ (only internal symmetries
possessing suitable \textit{real} representations can be
implemented on multiplets of Majorana spinors $\chi$). An
experimental indication of the viability of such nonabelian polar
decompositions might be the confirmation of \textit{Majorana} mass
terms in the lepton sector, suggesting the possibility of
admixtures with the \textit{Ur}-field $\chi$. Further
consideration of such generalisations is deferred to future work.

\subsubsection*{Acknowledgements}
The authors are grateful to Robert Delbourgo and also Simon
Wotherspoon for helpful comments.

\begin{appendix}

\section*{Appendix}
\renewcommand{\theequation}{\Alph{section}.\arabic{equation}}

\section{The Dirac algebra in $2+1$ dimensions.}

 The fundamental
relation defining the gamma matrices in 2+1 dimensions is
\begin{align}\label{anti}
\anticom{\gamma^{\mu}}{\gamma^{\nu}}=2\eta^{\mu\nu},
\end{align}
where the metric is
\begin{align}
\eta=\left( {\begin{array}{*{20}c}
   1 & 0 & 0  \\
   0 & { - 1} & 0  \\
   0 & 0 & { - 1}  \\
\end{array}} \right).\nonumber
\end{align}
The anti-commutation relation (\ref{anti}) can be satisfied by a
set of $2\times{2}$ matrices generated from the Pauli matrices
\begin{alignat}{3}
\sigma^1&=\left( {\begin{array}{*{20}c}
   0 & 1  \\
   1 & 0 \\
\end{array}} \right),\qquad
\sigma^2&=\left( {\begin{array}{*{20}c}
   0 & -i  \\
   i & 0 \\
\end{array}} \right),\qquad
\sigma^3&=\left( {\begin{array}{*{20}c}
   1 & 0  \\
   0 & { - 1} \\
\end{array}} \right).\nonumber
\end{alignat}
Any generic spinor in 2+1 dimensions has two components
\begin{align}
\psi=\left(\begin{array}{*{20}c}
   {u_1}  \\
   {u_2}  \\
\end{array}\right).\nonumber
\end{align}
These independent components can be used to construct independent
positive and negative energy solutions. For the purposes of
explicit calculation it is useful to take a particular
representation of the gamma matrices.  Using the Pauli matrices a
standard representation can be chosen as
\begin{alignat}{3}
\gamma^0&=\sigma^3,\qquad\gamma^1&=i\sigma^1,\qquad\gamma^2&=i\sigma^2.\nonumber
\end{alignat}
In odd dimensions there is no $\gamma^5$ matrix. The charge
conjugation matrix $C$ can be chosen to be,
\begin{align}
C&=-\gamma^2=-i\sigma^2,\nonumber\\
C^{-1}&=-C=C^{\intercal}=C^{\dagger}.\nonumber
\end{align}
A Majorana spinor then takes the form
\begin{align}\label{majcon}
\chi=\left(\begin{array}{*{20}c}
   {v}  \\
   {v^*}  \\
\end{array}\right).
\end{align}
Taking solutions of the Dirac equation as anti-commuting field
operators it follows that for Majorana fields
\begin{align}
\overline{\chi}\gamma^{\mu}\chi=\overline{\chi}\sigma^{\mu\nu}\chi=0,\nonumber
\end{align}
indicating that the Majorana field is truly non-interacting with
respect to the usual couplings.

The algebra is completed by the relations
\begin{alignat}{2}
\gamma^{\mu}\gamma^{\nu}&=\eta^{\mu\nu}-i\epsilon^{\mu\nu\lambda}\gamma_{\lambda},
\qquad\epsilon^{012}=1,\nonumber\\
\gamma^{\mu}\gamma^{\alpha}\gamma^{\nu}&=
\eta^{\mu\alpha}\gamma^{\nu}-\eta^{\mu\nu}\gamma^{\alpha}+\eta^{\alpha\nu}\gamma^{\mu}-i\epsilon^{\mu\alpha\nu}.\nonumber
\end{alignat}
The generators of Lorentz transformations are then given by
\begin{align}
\sigma^{\mu\nu}=\frac{i}{2}\com{\gamma^{\mu}}{\gamma^{\nu}}=\epsilon^{\mu\nu\lambda}\gamma_\lambda,\nonumber
\end{align}
and the spinor representation of the Lorentz group can be
expressed as
\begin{align}\label{spinrep}
S(\Lambda)&=e^{\frac{i}{4}\omega_{\mu\nu}\sigma^{\mu\nu}}=e^{\frac{i}{4}\theta_{\mu}\gamma^{\mu}}\nonumber\\
\theta^{\lambda}&=\epsilon^{\mu\nu\lambda}\sigma_{\mu\nu}.
\end{align}
As an example of polar decompositions, consider the free particle
solutions
\begin{alignat}{2}\label{plane}
\psi=N\left(\begin{array}{*{20}c}
  1 \\
   \frac{ip_+}{E+m} \\
\end{array}\right)e^{-ip_{\mu}x^{\mu}},\qquad\psi_c=N\left( \begin{array}{*{20}c}
   \frac{-ip_-}{E+m}\\
    1 \\
\end{array} \right)e^{ip_{\mu}x^{\mu}},
\end{alignat}
for $E=+\sqrt{p^2+m^2}$, $N$ is a normalization factor and
$p_+=p^1+ip^2=p_-^*$. Analogously free solutions in polar
coordinates, $(x,y)\rightarrow(r,\theta)$, can be defined in terms
of Bessel functions as
\begin{alignat}{2}\label{bessel}
\psi=N\left(\begin{array}{*{20}c}
  J_{\alpha}(pr) \\
   \frac{-pe^{i\theta}}{E+m}J_{\alpha+1}(pr) \\
\end{array}\right)e^{i(\alpha\theta-Et)},\qquad\psi_c=N\left( \begin{array}{*{20}c}
  \frac{-pe^{-i\theta}}{E+m}J_{\alpha+1}(pr)\\
    J_{\alpha}(pr) \\
\end{array} \right)e^{-i(\alpha\theta-Et)},
\end{alignat}
where $p=+\sqrt{E^2-m^2}$ and $J_{\alpha}(pr)$ are the Bessel
functions of the first kind, \cite{tranter}. From the plane wave
solutions (\ref{plane}) and using $\chi$ as in (\ref{majcon}) the
polar decomposition takes on the form

\begin{align}
v=\textstyle{\frac{1}{2}}N\left[{(E+m-ip_-)\cos{p_{\mu}x^{\mu}}-i(E+m+ip_-)\sin{p_{\mu}x^{\mu}}}\right]
\end{align}
\begin{alignat}{3}
n^0&=\frac{E}{m},\qquad {n}^1&=\frac{p^1}{m},\qquad
{n}^2&=\frac{p^2}{m}.
\end{alignat}
Whereas for the charge conjugate solution $\psi_c$
\begin{alignat}{3}
n^0_c&=\frac{-E}{m},\qquad {n}^1_c&=\frac{-p^1}{m},\qquad
{n}^2_c&=\frac{-p^2}{m}.
\end{alignat}
Thus the property $n_{\mu}\rightarrow{-}n_{\mu}$ under charge
conjugation is in keeping with what happens under charge conjugation
to the plane wave solution $\psi \rightarrow\psi_c$.

For the solutions (\ref{bessel}) we have
\begin{align}
v=&\textstyle{\frac{1}{2}}N\left[{e^{i(\alpha\theta-Et)}J_{\alpha}(pr)-\frac{pe^{-i\theta}}{E+m}J_{\alpha+1}(pr)e^{-i(\alpha\theta-Et)}}\right],\nonumber\\
j^0=&N^2[J_{\alpha}^2(pr)+\textstyle{\frac{p^2}{(E+m)^2}}J_{\alpha+1}^2(pr)]=j^0_c,\nonumber\\
j^1=&\frac{2N^2p}{E+m}J_{\alpha}J_{\alpha+1}\sin{\theta}=j^1_c,\nonumber\\
j^2=&\frac{-2N^2p}{E+m}J_{\alpha}J_{\alpha+1}\cos{\theta}=j^2_c,\nonumber\\
j^r=&\frac{2N^2p}{E+m}J_{\alpha}J_{\alpha+1}=j^r_c,\nonumber\\
j^{\theta}=&(\theta-\textstyle{\frac{\pi}{2}})=j^{\theta}_c,\nonumber\\
\overline{\psi}\psi=&N^2\left[J^2_{\alpha}(pr)-\textstyle{\frac{p^2}{(E+m)^2}J^2_{\alpha+1}(pr)}\right]=-\overline{\psi_c}\psi_c.
\end{align}
So again we have $n^{\mu}\rightarrow-n^{\mu}$ under charge
conjugation.
\end{appendix}

\end{document}